\documentclass[prl,12pt]{revtex4-2} \synctex=1 
\usepackage{graphicx}  
\usepackage{amsmath,amssymb,bm,bbm} 
\usepackage[colorlinks=true]{hyperref}  
\hypersetup{
    bookmarks=true,         
    unicode=false,          
    pdftoolbar=true,        
    pdfmenubar=true,        
    pdffitwindow=false,     
    pdfstartview={FitH},    
    pdftitle={},    
    pdfauthor={},     
    pdfsubject={},   
    pdfcreator={},   
    pdfproducer={}, 
    pdfkeywords={} {} {}, 
    pdfnewwindow=true,      
    colorlinks=true,       
    linkcolor=magenta, 
    citecolor=blue,        
    filecolor=magenta,      
    urlcolor=blue           
} 

\usepackage{pdfpages}
\usepackage{pgffor}
\makeatletter
\AtBeginDocument{\let\LS@rot\@undefined}
\makeatother

\begin{document}
\begin{center}
{\bf RVB and odd $\mathbb{Z}_2$ spin liquids}
\end{center}
\thispagestyle{empty}

This article contains the published version of \href{https://journals.jps.jp/page/jpsj/suppl}{ Journal of the Physical Society of Japan {\bf 69}, Suppl. B, 1 (2000)} which is not available online, or in many libraries. Here is a link to a \href{http://sachdev.physics.harvard.edu/p91.pdf}{scanned copy of the printed version}.

Also attached are scanned pages from my notebook \#24 written July 30, 1990. The content of these notes were outlined, and final results quoted, 
in \href{https://journals.aps.org/prb/abstract/10.1103/PhysRevB.44.686}{Phys. Rev. B {\bf 44}, 686 (1991)} and \href{https://journals.aps.org/prb/abstract/10.1103/PhysRevB.45.12377}{Phys. Rev. B {\bf 45}, 12377 (1992)}. The full contents of these notes were published in 
\href{https://journals.jps.jp/page/jpsj/suppl}{ Journal of the Physical Society of Japan {\bf 69}, Suppl. B, 1 (2000)}. I note that the results in \href{https://journals.jps.jp/page/jpsj/suppl}{ Journal of the Physical Society of Japan {\bf 69}, Suppl. B, 1 (2000)} did not appear elsewhere in the literature, between the 1990 note and the 2000 publication.

These publications and notes derive dual models of (what are now called) $\mathbb{Z}_2$ spin liquids, and show for the first time that $\mathbb{Z}_2$ spin liquids come in 2 varieties: these are now labeled `odd' and `even' $\mathbb{Z}_2$ spin liquids. For spin quantum antiferromagnets, odd (even) spin liquids appear for half-integer (even-integer) spin per unit cell. 

The even case corresponds ultimately to the $\mathbb{Z}_2$ gauge theory studied by 
Wegner (and later in the toric-code model by Kitaev), and this can have a trivial phase with no broken symmetry. 

The odd case is new, and was considered here for the first time. In this case, the $\mathbb{Z}_2$ gauge theory has a background gauge charge, and the confining phase is shown here to break lattice symmetry by the appearance of valence bond solid order. This ensures consistency with the Lieb-Schultz-Mattis-Oshikawa-Hastings theorems. It is also shown that vison (or `$m$' particle) excitations are at least doubly degenerate for the odd $\mathbb{Z}_2$ spin liquid. 

Anderson's resonating valence bond (RVB) theory was for Mott insulators with one electron per site. Such insulators must obey the LSMOH theorem, and cannot have a trivial phase. So Wegner's $\mathbb{Z}_2$ gauge theory is not a correct effective theory for the RVB phase. Instead, it is the {\it odd\/}  $\mathbb{Z}_2$ gauge theory which is the correct effective theory for the RVB phase, a claim that appeared first in \href{https://journals.aps.org/prb/abstract/10.1103/PhysRevB.44.686}{Phys. Rev. B {\bf 44}, 686 (1991)}. The `odd' nature of 
the effective theory has many important physical consequences, including changes in the universality classes of transitions out of the RVB phase. \\
\begin{flushright}
Subir Sachdev
\end{flushright}
\newpage
\title
{
Translational symmetry breaking in \\ two-dimensional
antiferromagnets and superconductors
}

\author
{
Subir Sachdev and Matthias Vojta
}

\affiliation
{
Department of Physics, Yale University, \\
P.O. Box 208120, New Haven, CT 06520-8120, USA
}

\date
{
October 6, 1999
}

\begin{abstract}
It was argued many years ago that translational symmetry
breaking due to the appearance of spin-Peierls ordering
(or bond-charge stripe order) is a fundamental property of the quantum
paramagnetic states of a large class of square lattice
antiferromagnets. Recently, such states were shown
to be a convenient point of departure for studying translational
symmetry breaking in doped antiferromagnets: these results are
briefly reviewed here with an emphasis on experimental implications.
In the presence of stronger frustration, it was also argued
that the insulating antiferromagnet can undergo a transition to a
deconfined state with no lattice symmetry breaking. This
transition is described by a fully-frustrated Ising model in a
transverse field: details of this earlier derivation of the Ising
model are provided here---this is motivated by the
reappearance of the same Ising model in a recent study of the
competition between antiferromagnetism and d-wave
superconductivity by Senthil and Fisher.
\end{abstract}
\maketitle
\setcounter{page}{1}

~\\
\begin{center}
\tt
Proceedings of the International Workshop on {\bf Magnetic
Excitations in Strongly Correlated Electrons}, Hamamatsu, Japan,
August 19-22, 1999.\\ To appear in the Journal of the Physical
Society of Japan.
\end{center}

\section{Introduction}

Following the fundamental neutron scattering observations of
Tranquada\cite{jtran}, there has been a vigorous resurgence\cite{birg,imai}
in the study of
states of two-dimensional strongly correlated systems
which break various types of translational symmetries
(to be precise: we will consider translational symmetry to be broken
when an observable invariant under spin rotations (and
which does not change the net charge of the system)
 does not
remain invariant under the space-group of the lattice).
In an early discussion of translational symmetry breaking in
two-dimensional antiferromagnets\cite{rsprl1,rsprb1}
it was argued that any quantum paramagnet accessed
by a continuous quantum transition from a N\'{e}el state must
have (for $S=1/2$ spins)
({\em i}) broken translational symmetry due to spin-Peierls
(or {\em bond}-charge density)
ordering, and ({\em ii}) confinement of spin-1/2 excitations
resulting in integer spin quasiparticle modes.
Other early work\cite{zaanen,schulz} used Hartree-Fock theory to
study the formation of {\em site}-charge density order in lightly
doped antiferromagnets. In this paper, we shall review
recent work\cite{vojta,lt}
in which these two apparently divergent physical effects were
studied in a combined footing---we shall argue that such an
analysis leads to considerable physical insight and has
significant experimental implications.

This paper is organized as follows. In Section~\ref{ins} we will
consider translational symmetry breaking in the quantum paramagnetic phases
of insulating antiferromagnets---this will be done using lattice
`height' and Ising models which are obtained after a duality
transformation. In
Section~\ref{dope} we will turn to the recent results\cite{vojta,lt}
on doped antiferromagnets, and their
implication for experiments.

\section{Insulating antiferromagnets}
\label{ins}

A convenient and physically transparent framework for discussing
the physics of the low-lying singlet excitations of a quantum
paramagnet is the ``quantum dimer model''\cite{dimer}. At first glance, the
formulation of the dimer model does appear rather arbitrary, with a
number of ad hoc assumptions which obscure the connection to
the Heisenberg antiferromagnet---instead, the dimer model is best viewed as
a caricature which captures some essential pieces of the physics.
We shall mainly use the dimer model
as an intuitive intermediate step towards introducing
dual `height' and Ising models. These dual models have also been
derived by a number of other methods, including systematic
semiclassical and large $N$ expansions\cite{rsprl1,rsprb1},
and these provide the more
formal justifications for the arguments here.
We will describe duality transformations
on an
effective model of resonating singlet bonds which show that
unfrustrated antiferromagnets are described by an effective height
model in 2+1 dimensions\cite{fradkiv,rsprb1,js1};
in the presence of frustration, the
height model gets modified to a fully frustrated, two-dimensional
Ising model in a
transverse field\cite{js2}. All of the results in this section were derived
and published some time ago\cite{fradkiv,rsprb1,js1,js2} ;
the frustrated Ising
model was derived by one of us\cite{ss} and the main
results were outlined in Ref.\cite{js2} , but details
have not been previously published. The presentation of the details
here was
motivated by the reappearance of the same Ising model in a
recent study of the competition between antiferromagnetism and
d-wave superconductivity by Senthil and Fisher\cite{senthil}.
We also note another recent study of lattice gauge theory models
of two-dimensional antiferromagnets by Nagaosa and
Lee\cite{nagaosa}; they examine theories related to those
considered in Ref\cite{js1} and in the present paper, but their
actions do not contain the crucial Berry phase terms
(as in (\ref{f10a}), (\ref{f10}), and (\ref{g10}))
associated
with the non-zero net matter density on each site even in the
undoped antiferromagnet---it is these Berry phase terms which are responsible
for frustration in the dual height or Ising spin representation,
and we believe their incorporation is essential for a proper
description of the physics.

\begin{figure}[t]
\centering
\includegraphics[width=4in]{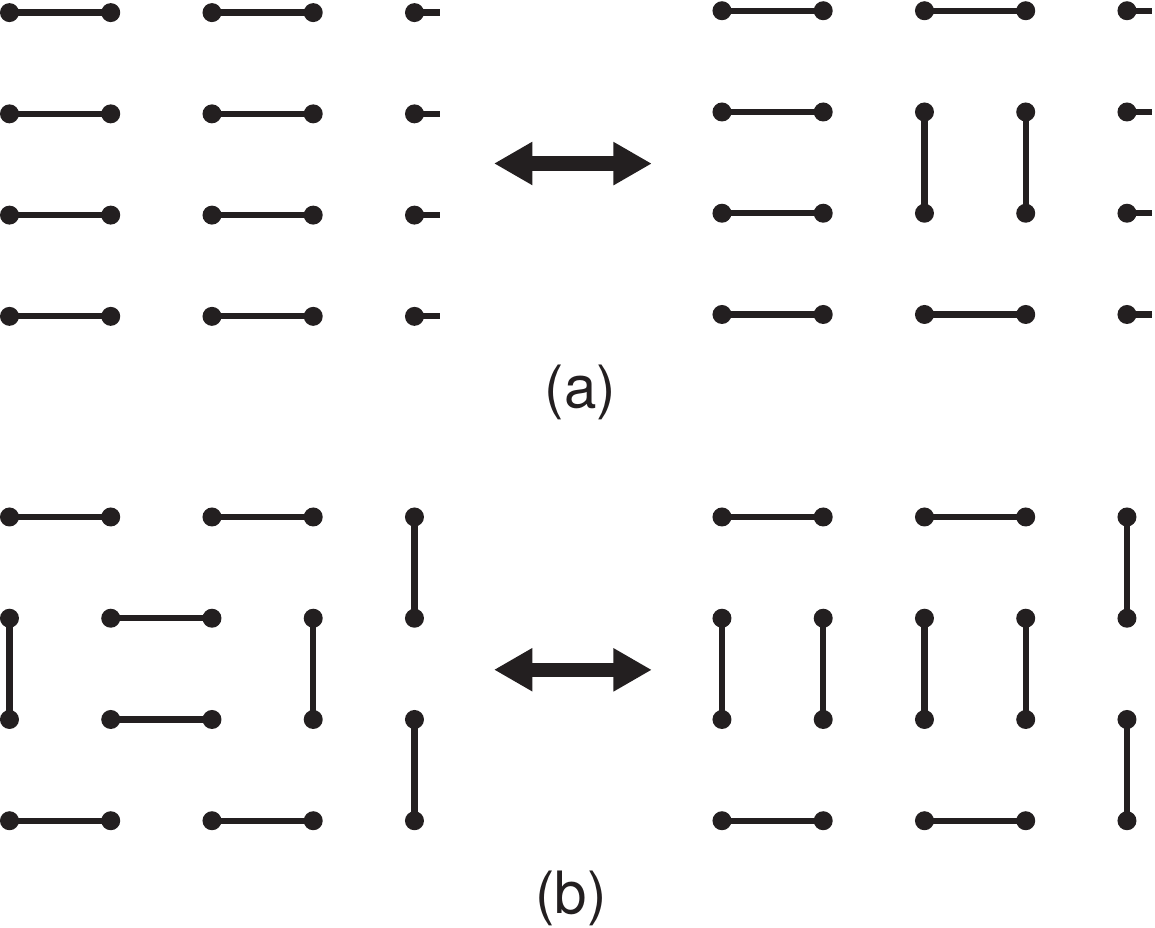}
\caption{Dimer coverings of the square lattice, each representing
a distinct state in the singlet Hilbert space. The two states
in (a) can resonate with each other as can the two in (b).
The columnar state in (a) has the maximum number of resonating
partners, and this drives the translational symmetry breaking.
}
\label{fig1}
\end{figure}

The central assertion of the dimer model is that it is appropriate
identify the Hilbert space of the low-lying singlet excitations of
the quantum paramagnet on a square lattice
with a set of dimer coverings\cite{dimer}. Each dimer
covering corresponds to different singlet state, and off-diagonal
matrix elements between such states therefore lead to a
``resonance'' between the dimers. Such resonance terms must
preserve the constraint of one and exactly one dimer terminating
at every site, and this severely restricts their form. Two
possible resonating terms are illustrated in Fig~\ref{fig1}.
One might now anticipate that the ground state of such a
resonating-valence-bond Hamiltonian is a liquid-like
superposition of all dimer coverings, but this turns out not to be
the case. The ground state in fact breaks lattice translational
symmetry\cite{rsprl1,rsprb1} {\em e.g.} for some range of parameters it looks like the
columnar ordered state at the top of Fig~\ref{fig1}. There is a
simple intuitive reason for this: the columnar state has the
maximum number of plaquettes about which the dimers are able to
resonate and still preserve the local constraints. This resonance
lowers its energy, and this effect is strong enough that the
system always picks out one of the four columnar states and
retains memory of its orientation. The ordering is therefore due
to a quantum ``order-from-disorder'' effect.

Let us now make these arguments more precise.
We will divide our discussion into two subsections. Section~\ref{bm} will
consider the model with dimers connecting nearest neighbors on the
square lattice. More generally, the results here will apply to
bipartite lattices in which the dimers only connect sites on
opposite sublattices. We will show that such models here are
described by effective height models in 2+1 dimensions\cite{rsprb1,js1}.
Section~\ref{nb} will consider models in which the dimers are
allowed to connect sites on the same sublattice---the effective
model for these will be the fully frustrated Ising model in a
transverse field\cite{js2}.

\subsection{Bipartite dimer models}
\label{bm}

We will only discuss the case of the nearest-neighbor, square
lattice dimer model here. We identify the dimers by site, $i$, of
the square lattice on their lower or left end. Let $\eta_i \hat{E}_{i\alpha}$
be the number operator for the dimer on site $i$ oriented in the $\alpha$
direction ($\alpha = x, y$); $\eta_i$ indicates the sublattice of
site $i$, and equals $+1$ on one sublattice and $-1$ on the other.
We have so far considered only a $S=1/2$ antiferromagnet, but
for general $S$ there will be exactly $2 S$ dimers emerging from
every site---this constraint translates to
\begin{equation}
\Delta_{\alpha} \hat{E}_{i \alpha} = 2 S \eta_i ,
\label{f1}
\end{equation}
where $\Delta_{\alpha}$ is the discrete lattice derivative in the $\alpha$
direction. The factors of $\eta_i$ where introduced so that the constraint
would have the Gauss-law form in (\ref{f1}), and the operator $\hat{E}_{i\alpha}$
is seen to be the analog of the electric field\cite{fradkiv}.
We also introduce an angular phase variable, $\hat{A}_{i
\alpha}$ (the analog of a compact $U(1)$ gauge field),
on every link which is canonically conjugate to $\hat{E}_{i
\alpha}$:
\begin{equation}
[\hat{A}_{i \alpha}, \hat{E}_{j \beta} ] = i \delta_{ij}
\delta_{\alpha \beta},
\label{f2}
\end{equation}
where it should be clear from the context when we mean $i =
\sqrt{-1}$,
and when $i$ is a site label.

We can now write down the Hamiltonian of the dimer model:
\begin{equation}
H_d = \frac{K_1}{2} \sum_{i,\alpha} \hat{E}_{i \alpha}^2
- K_2 \sum_i \cos(\epsilon_{\alpha\beta} \Delta_{\alpha}
\hat{A}_{i \beta}).
\label{f3}
\end{equation}
The first term, proportional to $K_1$ is only non-trivial when $2S > 1$, and it ensures
that the density of dimers is as uniform as possible. It follows
from the commutation relations (\ref{f2}) that the second term,
proportional to $K_2$, flips dimers around a plaquette and so
induces the `resonance' shown in Fig~\ref{fig1}.

We will now write down a path integral representation of the
partition function of $H_d$ by following a standard route. We
insert complete sets of $\hat{E}_{i \alpha}$ eigenstates at small
imaginary time intervals $\Delta \tau$. The matrix elements of the
cosine term in $H_d$ are evaluated by replacing it with the
Villain periodic Gaussian form (this is the only `approximation' made in our
duality mappings - all other transformations below are exact):
\begin{equation}
\exp\left(K_2 \Delta \tau \cos(\epsilon_{\alpha\beta} \Delta_{\alpha}
\hat{A}_{i \beta})\right) \rightarrow \sum_{B_a} \exp \left(
- \frac{B_a^2}{2 K_2 \Delta \tau} + i B_a \epsilon_{\alpha\beta} \Delta_{\alpha}
\hat{A}_{i \beta} \right).
\label{f4}
\end{equation}
Here $B_a$ is an integer-valued field on the sites, $a$, of the
dual lattice; there is an obvious relationship between the
location of the direct lattice site, $i$, and the dual lattice
site $a$, but we will not specify it explicitly to avoid cluttering up the
notation (we will consistently use the labels $i,j\ldots$ for sites
on the direct lattice, and the labels $a,b\ldots$ for sites on the dual lattice).
In the present case, $a$ resides at the center of
plaquette around which the `flux' $\epsilon_{\alpha\beta} \Delta_{\alpha}
\hat{A}_{i \beta}$ resonates the dimers. It is also convenient to
introduce a three-vector notation in space time: we define the
integer-valued `electromagnetic flux' vector
$F_{a\mu} = (E_{iy}, -E_{ix}, -B_a)$ on the dual lattice sites, where
the index $\mu = (x,y,\tau)$ (we will consistently use the labels
$\alpha,\beta \ldots$
to represent spatial components only, while $\mu,\nu,\lambda\ldots$ will represent
three-dimensional spacetime components). Here $E_{i \alpha}$ refer to the
integer eigenvalues of the operator $\hat{E}_{i \alpha}$ which are summed
over in each time step. Now the
partition function of $H_d$ can be written in the compact form
\begin{equation}
Z_F = \sum_{\{F_{a \mu}\}} \exp \left( - \frac{e^2}{2}
\sum_{a,\mu} F_{a \mu}^2 \right) \prod_{a,\mu} \delta \left(
\epsilon_{\mu\nu\lambda}
\Delta_{\nu} F_{a \lambda} - 2S \eta_i \delta_{\mu\tau} \right)
\label{f5}
\end{equation}
Here the sum is over the integer-valued field $F_{a \mu}$ which
resides on the sites of the dual cubic lattice in spacetime; the
delta function constraint imposes `Gauss's law' (\ref{f1}) and the
`Maxwell equation' between $E$ and $B$ which follows from the
matrix elements of (\ref{f4}). The time spacing $\Delta \tau$ has
been chosen so that the coupling $e^2 = K_1 \Delta \tau  = 1/(K_2 \Delta
\tau)$.

\begin{figure}
\centering
\includegraphics[width=5.5in]{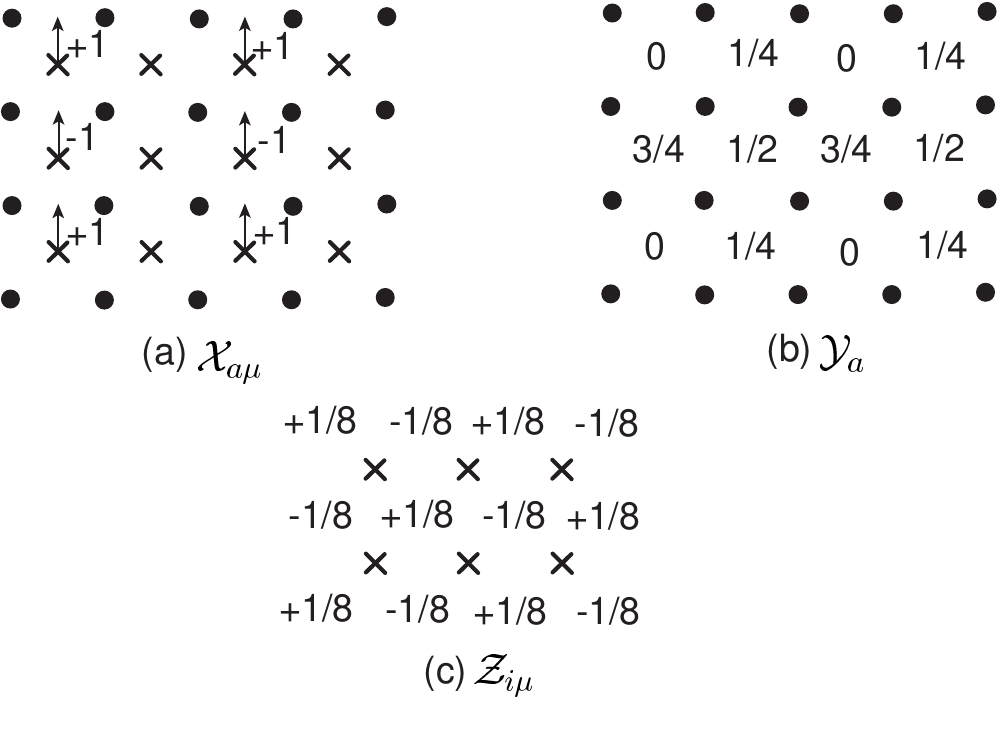}
\caption{The values of the only non-zero components of the fixed
field ${\cal X}_{a \mu}$, ${\cal Y}_a$, and ${\cal Z}_{i \mu}$.
The circles (crosses) are the sites of the direct (dual) lattice.
In (c), only the $\mu=\tau$ component of ${\cal Z}_{i \mu}$ is
non-zero and its values are shown.
}
\label{fig2}
\end{figure}

We now solve the constraint in (\ref{f5}) by writing $F_{a\mu}$ as
the sum of a particular solution and the general solution of the
homogeneous equation:
\begin{equation}
F_{a \mu} = \Delta_{\mu} N_a + 2 S {\cal X}_{a \mu}.
\label{f6}
\end{equation}
Here $N_a$ is a fluctuating integer-valued field on the dual lattice
sites, while ${\cal X}_{a \mu}$ is a {\em fixed} field independent of $\tau$
satisfying
\begin{equation}
\epsilon_{\mu\nu\lambda} \Delta_{\nu} {\cal X}_{a \lambda} = \eta_i
\delta_{\mu \tau}.
\label{curlx}
\end{equation}
A convenient
choice is to take ${\cal X}_{a x} = 0$, ${\cal X}_{a \tau} = 0$,
and ${\cal X}_{a y}$
as shown in Fig~\ref{fig2}a, taking the values $\pm 1$ on every second column
of sites and zero otherwise.
For future manipulations, it is convenient to split ${\cal X}_{a \mu}$
into curl-free and divergence-free parts by writing
\begin{equation}
{\cal X}_{a \mu} = \Delta_{\mu} {\cal Y}_a + \epsilon_{\mu \nu
\lambda} \Delta_{\nu} {\cal Z}_{i \lambda},
\label{f7}
\end{equation}
where again ${\cal Y}_a$ and ${\cal Z}_{i \mu}$ are fixed fields
independent of $\tau$ and their values are shown in
Fig~\ref{fig2}b,c; ${\cal Y}_a$ takes the values
$0,1/4,1/2,3/4$ on the four dual sublattices, while
${\cal Z}_{i \mu} = \delta_{\mu \tau} \eta_i /8$.

We have now assembled all the ingredients necessary for the final
representations of the partition function. There are three
different formulations, and each present a valuable physical
perspective and can be used for further quantitative analysis.
We will present them in turn in the following subsections.

\subsubsection{Height model}
\begin{figure}
\centering
\includegraphics[width=3in]{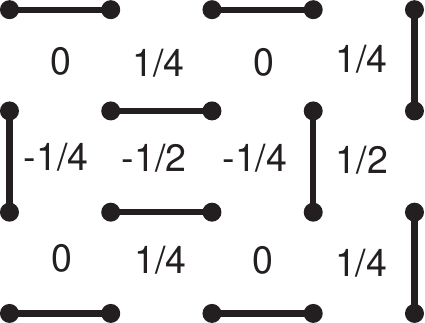}
\caption{A set of values of $H_a$ satisfying (\protect\ref{e2})
and the corresponding
dimer covering.
}
\label{fig3}
\end{figure}

We
insert (\ref{f6}) into (\ref{f5}) and obtain the height model
partition function
\begin{equation}
Z_H = \sum_{\{N_a\}} \exp \left( - \frac{e^2}{2} \sum_{a,\mu}
\left( \Delta_{\mu} N_a + 2 S {\cal X}_{a \mu} \right)^2 \right).
\label{e3a}
\end{equation}
For the purposes of this section, it is convenient to insert
the decomposition (\ref{f7}) in (\ref{e3a}) and
obtain\cite{rsprl1,rsprb1,js1}
\begin{equation}
Z_H^{\prime} = \sum_{\{N_a\}} \exp \left( - \frac{e^2}{2} \sum_{a,\mu}
\left( \Delta_{\mu} H_a \right)^2 \right),
\label{e3}
\end{equation}
where the `heights' $H_a$ are defined by
\begin{equation}
H_a = N_a + 2S {\cal Y}_a.
\label{defh}
\end{equation}
Notice that ${\cal Z}_a$ has dropped out--this is a general
property of all representations of bipartite dimer models,
but the non-bipartite models in
Section~\ref{nb} will depend upon ${\cal Z}_a$.
We can view $H_a$ as the heights of a 3-dimensional interface
which are restricted to take values equal (for $S=1/2$) to those in
Fig~\ref{fig2}b plus arbitrary integers. It is instructive to note
here, for $S=1/2$, a direct connection between these heights and the original
dimer model we began with\cite{zheng}. For large $e^2$, the partition function
$Z_H^{\prime}$ will be dominated by configurations in which neighboring
heights are as close to each other as possible; the manifold of
minimum action is defined by heights which satisfy
\begin{equation}
|H_a - H_b | < 1 ~~\mbox{for every pair of nearest neighbors
$a,b$},
\label{e2}
\end{equation}
If we consider two height configurations to be equivalent if they
differ only by a constant integer shift $H_a \rightarrow H_a + p$,
then the set of all such equivalence
classes is precisely equivalent to the set of dimer coverings.
The identification is illustrated in Fig~\ref{fig3}: neighboring
heights differ either by large (3/4) or small (1/4) steps, and we
place a dimer on every link intersecting a large step.

\subsubsection{sine-Gordon model}
Promote
the integer valued field $N_a$ to a real-valued field $\varphi_a$
by the Poisson summation formula. After shifting the real field
by $\varphi_a \rightarrow \varphi_a - {\cal Y}_a$ we obtain
from (\ref{f5},\ref{f6},\ref{f7})
\begin{equation}
Z_{sG} = \prod_a \int_{-\infty}^{\infty} d \varphi_a \sum_{\{P_a\}}
\exp \left(- \frac{e^2}{2} \sum_{a,\mu}
\left( \Delta_{\mu} \varphi_a \right)^2 + 2 \pi i \sum_a P_a \left(
\varphi_a - {\cal Y}_a \right) \right).
\label{e4}
\end{equation}
If we now truncate the summation over $P_a$ by adding a fugacity
$e^{(\ln y) P_a^2}$, then
$\sum_{p_a} e^{2\pi i P_a (\varphi_a - {\cal Y}_a) + (\ln y) P_a^2} \approx
e^{2 y \cos(2 \pi (\varphi_a - {\cal Y}_a))}$, and $Z_{sG}$ is seen to be
a version of the sine-Gordon field theory in 2+1 dimensions\cite{rsprl1,rsprb1}.

\subsubsection{Compact QED}
\label{cqed}
Define $M_{a \mu} = \Delta_{\mu} N_a$; so $M_{a \mu}$ is an integer-valued field
satisfying the constraints
\begin{equation}
\epsilon_{\mu\nu\lambda} \Delta_{\nu} M_{a \lambda} = 0.
\label{f8}
\end{equation}
We impose these constraints by a real Lagrange multiplier field ${A_{i
\mu}}$ by inserting factors of
\begin{equation}
\int_{-\pi}^{\pi} \frac{d A_{i \mu}}{2 \pi} \exp \left(
i A_{i \mu} \epsilon_{\mu\nu\lambda} \Delta_{\nu} M_{a \lambda}
\right)
\label{f9}
\end{equation}
in $Z_H$ in (\ref{e3a}). The sum over the $M_{a \mu}$ can now be performed using
the Poisson summation formula and we obtain after using
(\ref{curlx})
\begin{equation}
Z_{cQED}= \sum_{\{Q_{a\mu}\}} \prod_{i,\mu}
\int_{-\pi}^{\pi} \frac{d A_{i \mu}}{2 \pi}
\exp \left(
- \frac{1}{2 e^2} \sum_{i,\mu} \left( \epsilon_{\mu\nu\lambda} \Delta_{\nu}
A_{i \lambda} - 2 \pi Q_{a \mu} \right)^2
+ i 2S \sum_i \eta_i A_{i\tau} \right),
\label{f10a}
\end{equation}
where the $Q_{a \mu}$ extend over all integers.
$Z_{cQED}$ is easily seen to be the Villain form of the partition
function of a compact $U(1)$ gauge theory in 2+1 dimensions.
There is an additional Berry phase carried by the world lines
of the background charge density of $2S$ particles on each site.
An alternative, and exactly equivalent, form of $Z_{cQED}$ can be
obtained by applying the above transformation to $Z_H^{\prime}$
(in (\ref{e3})) rather than to $Z_H$. Then, exactly the same steps
lead to
\begin{equation}
Z_{cQED}^{\prime} = \sum_{\{Q_{a\mu}\}} \prod_{i,\mu}
\int_{-\pi}^{\pi} \frac{d A_{i \mu}}{2 \pi} \exp \left(
- \frac{1}{2 e^2} \sum_i \left( \epsilon_{\mu\nu\lambda} \Delta_{\nu}
A_{i \lambda} - 2 \pi Q_{a \mu} \right)^2 +
i 4 \pi S \sum_a {\cal Y}_a \Delta_{\mu} Q_{a \mu} \right)
\label{f10}
\end{equation}
The instanton number of the gauge theory is $\Delta_{\mu} Q_{a
\mu}$, and to the Berry phase is now carried by the instantons:
each instanton has a Berry phase of $e^{i 4 \pi
S {\cal Y}_a}$. It is a remarkable fact that precisely the {\em same
Berry phase} was obtained in very different semiclassical and large
$N$ analysis of bipartite antiferromagnets\cite{haldane,rsprb1}.
It is also worth reiterating that $Z_{cQED}$ (in which the Berry phase
is carried by the world lines of the $2S$ particles on each site)
and $Z_{cQED}^{\prime}$ (in which the Berry phase is associated
with the instantons) are exactly equivalent\cite{js1}---they merely do the
book-keeping of the total Berry phase somewhat differently.

\begin{figure}
\centering
\includegraphics[width=3in]{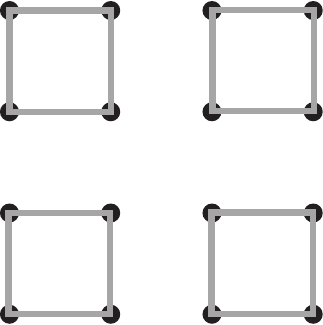}
\caption{Another possible ground state of the dimer model: the
plaquette state.
}
\label{fig4}
\end{figure}

The properties of the theories $Z_H^{\prime}$, $Z_{sG}$, and $Z_{cQED}^{\prime}$
are
reasonably well understood and have been described elsewhere\cite{rsprb1,zheng,js1}.
The height model
is in a smooth phase phase for all values of $e^2$ (there
is no roughening transition), while the compact $U(1)$ gauge
theory is always confining. In terms of the underlying dimer model, this
means that translational symmetry is spontaneously broken and the
dimers crystallize in a bond-charge ordered state. A likely
configuration is the columnar state in Fig~\ref{fig1}a, but it is
not ruled out that one of these theories could exhibit a
transition to some other confining state in which the details of
the lattice symmetry breaking are different: another state\cite{cavo} which
occurs in the parameter space of $Z_{sG}$ is the plaquette state
of Fig~\ref{fig4}.

\subsection{Non-bipartite dimer models}
\label{nb}

We will now relax the constraint that the dimers must connect
sites on opposite sublattices. We will show that
such a step reduces the height
model to a 2+1 dimensional fully frustrated Ising model in a
transverse field, following the derivation\cite{ss}
outlined in Ref\cite{js2}.

\begin{figure}
\centering
\includegraphics[width=3.5in]{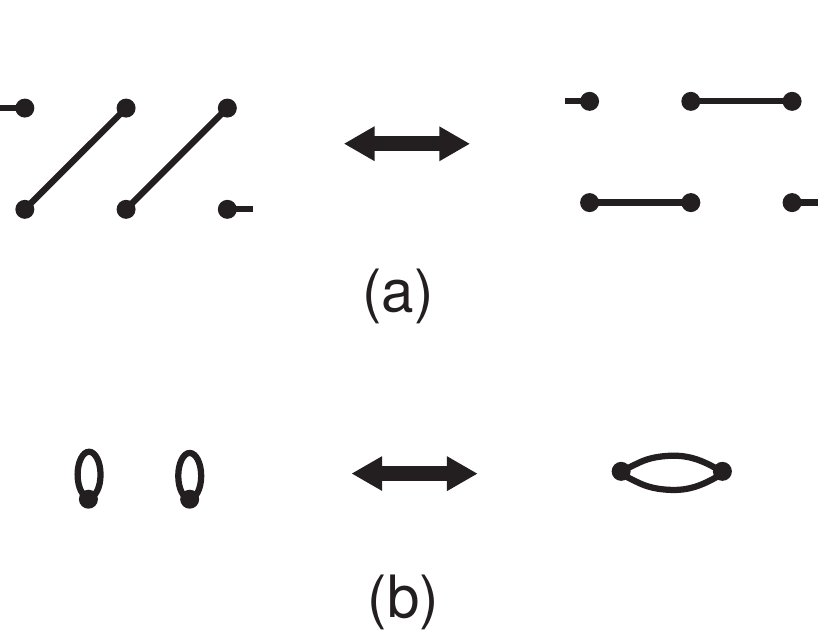}
\caption{({\em a}) Dimers which connect sites on the same sublattice
resonating with dimers connecting sites on opposite sublattices.
({\em b}) The same sublattice dimers have now been collapsed onto
monomers representing singlet bonds between electrons on the same
site. We argue that the deformation from ({\em a}) to ({\em b})
should have no significant effect on the long distance properties:
it is crucial to keep track of which sublattice the two ends of
a singlet bond reside, and local deformations which preserve this
information will not modify the long distance physics.
The diagonal dimers in (a) and the monomers in (b) both carry net
charge $\pm 2$, and this is the key factor which controls their
effect on the long distance physics.
}
\label{fig5}
\end{figure}

An example of a dimer state connecting sites on the same
sublattice is
shown in Fig~\ref{fig5}a. Notice that a pair of such `diagonal'
dimers can resonate with a pair of dimers connecting opposite
sublattices.
However models
which include terms like those in Fig~\ref{fig5}a
do not appear to be amenable to the {\em exact}
duality transformations we have considered in Section~\ref{bm}.
So we have two alternatives for further analysis: we could be
satisfied by approximate duality transformations on models which
include the resonance in Fig~\ref{fig5}a, or we could modify the
underlying dimer model to allow duality to proceed smoothly.
Both alternatives lead to essentially the same final result, but
we will choose the latter for presentation here as it more clearly
isolates the approximation made.
The modification of the dimer model is aided by
the physical picture developed in
the large-$N$ theories of the effects of
frustration presented in Refs \cite{rsprl2,srijmp}.
The key property of the diagonal dimers in Fig~\ref{fig5}a
is that they connect sites on the same sublattice, and such dimers
carry net charge $\pm 2$ under the compact QED representation of
Section~\ref{cqed}. This will be shown explicitly in more detail
below, but can be seen quickly as follows. From the representation
(\ref{f10a}) we see that the model behaves as if there are
background charges of $+2S$ on one sublattice, and charges of $-2S$
on the other sublattice. A dimer connecting sites on opposite
sublattices absorbs (in the Gauss Law constraint (\ref{f1}))
charges of $\pm 1$ on its two ends; so such a dimer has net charge
0. However we do need a vector potential for a gauge invariant
connection between the charges on the two ends, and this is why
the number operator of the dimer $\sim e^{i A_{i\alpha}}$ (see
(\ref{f2})). Let us now apply the same argument to the diagonal
dimers of Fig\ref{fig5}a. These connect sites on the same
sublattice and therefore carry net charge $\pm 2$; however, we also
need a vector potential to connect such spatially separated
charges, and so the number operator of a diagonal operator
must
also contain the necessary path integral of $A_{i \alpha}$. It is
this combined role of diagonal dimers that precludes a
straightforward
duality transformation, but it also suggests a simplifying
modification: the important thing is that the dimer carry charge $\pm 2$
while preserving the fact that each site has net charge $\pm 2 S$;
so we collapse the two ends of the diagonal dimers onto the same
site, as in Fig~\ref{fig5}b. The resulting {\em monomer} trivially satisfies the
requirement of representing a singlet bond between electrons on
the same sublattice and thus carries charge $\pm 2$; moreover, no
vector potential is required as the charges on its two ends are on
the same spatial location. Further, such monomers can resonate,
like the diagonal dimers,
between dimers connecting sites on opposite sublattices, as shown
in Fig~\ref{fig5}.
Of course such a model requires that the
number of singlet bonds allowed to emerge from a site be at least
two {\em i.e.} $2S \geq 2$. Nevertheless, we maintain that generic
properties of the model studied here apply also to $S=1/2$ models
with configurations such as those in Fig~\ref{fig5}a.

Let us now write down the Hamiltonian of the monomer-dimer model
containing the resonance in Fig~\ref{fig5}b.
Let $\eta_i \hat{n}_i$ be the monomer number operator on site $i$.
Then the constraint that there are exactly $2S$ singlet bonds
emerging from site $i$ modifies the `Gauss Law' constraint (\ref{f1})
to
\begin{equation}
\Delta_{\alpha} \hat{E}_{i \alpha} + 2 \hat{n}_i  = 2 S \eta_i.
\label{g1}
\end{equation}
This equation shows that in addition to the background `charge' of
$2S$, there is\cite{rsprl2,srijmp}, as claimed above,
a fluctuating charge density of a `Higgs field' of monomers of charge 2.
We also introduce a angular phase variable $\hat{\phi}_i$
canonically conjugate to $\hat{n}_i$:
\begin{equation}
[\hat{\phi}_i, \hat{n}_j] = i \delta_{ij}
\label{g2}
\end{equation}
We can easily now extend the dimer Hamiltonian $H_d$ in (\ref{f3}) to
include the resonance in Fig~\ref{fig5}b, induced by the following cosine term
proportional to $K_4$:
\begin{equation}
\widetilde{H}_d = \frac{K_1}{2} \sum_{i,\alpha} \hat{E}_{i \alpha}^2
+ \frac{K_3}{2} \sum_i \hat{n}_i^2
- K_2 \sum_i \cos(\epsilon_{\alpha\beta} \Delta_{\alpha}
\hat{A}_{i \beta})
-K_4 \sum_{i,\alpha} \cos( \Delta_{\alpha} \hat{\phi}_i - 2 \hat{A}_{i \alpha} ).
\label{g3}
\end{equation}
For convenience we have also added an additional term,
proportional to $K_3$, which controls the density of monomer
fluctuations.

The remainder of this subsection describes the analysis of the Hamiltonian
$\widetilde{H}_d$, with the quantum dynamics defined by the commutation relations (\ref{f2})
and (\ref{g2}), while preserving the local constraint (\ref{g1}).
The procedure will closely parallel that followed in
Section~\ref{bm}.
We insert complete sets of $\hat{E}_{i \alpha}$ and $\hat{n}_{i}$
eigenstates at intervals of imaginary time $\Delta \tau$ in the
partition function.
As in (\ref{f4}), we replace the exponential of the new cosine
term proportional to $K_4$ by a Villain form but with a decoupling
field $k_{i \alpha}$. After evaluating the matrix elements between
the complete sets of states, defining the three-vector integer-valued `current'
$j_{i \mu} = (k_{ix}, k_{iy}, -n_i)$ we find that $Z_F$ in
(\ref{f5}) is replaced by
\begin{equation}
\widetilde{Z}_F = \sum_{\{F_{a \mu}\}, \{j_{i\mu} \}} \exp \left( - \frac{e^2}{2}
\sum_{a,\mu} F_{a \mu}^2 - \frac{g}{2}
\sum_{i, \mu} j_{i \mu}^2 \right) \prod_{a,\mu} \delta \left( \epsilon_{\mu\nu\lambda}
\Delta_{\nu} F_{a \lambda}+ 2j_{i\mu} - 2S \eta_i \delta_{\mu\tau} \right)
\prod_{i} \delta( \Delta_{\mu} j_{i \mu} ),
\label{g5}
\end{equation}
where, for simplicity, we have chosen the coupling $g = K_3 \Delta \tau =
1/(K_4 \Delta \tau)$. The sum in (\ref{g5}) is over integer valued
fields $F_{a \mu}$, on the dual lattice, and $j_{i \mu}$ on the
direct lattice. The solution of the constraints analogous
to (\ref{f6}) is
\begin{eqnarray}
j_{i\mu} &=& \epsilon_{\mu\nu\lambda} \Delta_{\nu} a_{a \lambda}
\nonumber \\
F_{a \mu} &=& \Delta_{\mu} N_a -2 a_{a \mu} + 2 S {\cal X}_{a
\mu},
\label{g6}
\end{eqnarray}
where $a_{a \mu}$ is an integer-valued `gauge' field on the links
of the dual lattice, while, as before, $N_a$ is an integer-valued
height field on the sites of the dual lattice. Inserting
(\ref{g6}) into (\ref{g5}) we obtain the generalization of the
height model (\ref{e3a})
\begin{equation}
\widetilde{Z}_H = \sum_{\{N_a\},\{a_{a\mu}\}} \exp \left( - \frac{e^2}{2} \sum_{a,\mu}
\left( \Delta_{\mu} N_a -2 a_{a \mu} + 2 S {\cal X}_{a \mu} \right)^2
- \frac{g}{2} \sum_{a,\mu} \left( \epsilon_{\mu\nu\lambda} \Delta_{\nu} a_{a\lambda}
\right)^2 \right).
\label{g7}
\end{equation}
We will refer to this partition function as the `gauged height
model'; the form (\ref{g7}) is the most convenient for subsequent
analysis and is one of the key expressions of this paper.
We also note an alternative form of
$\widetilde{Z}_H$, analogous to (\ref{e3}), obtained by inserting
(\ref{f7}) into (\ref{g7})
\begin{equation}
\widetilde{Z}_H^{\prime}
= \sum_{\{N_a\},\{a_{a\mu}\}} \exp \left( - \frac{e^2}{2} \sum_{a,\mu}
\left( \Delta_{\mu} H_a -2 a_{a \mu}  \right)^2
- \frac{g}{2} \sum_{a,\mu} \left( \epsilon_{\mu\nu\lambda} \Delta_{\nu} a_{a\lambda}
- \frac{2S e^2}{g} {\cal Z}_{i\mu} \right)^2 \right),
\label{g8}
\end{equation}
where the heights ${H}_a$ are defined in (\ref{defh}).
As promised earlier, ${\cal Z}_{i \mu}$ does not drop out of the
models of this subsection.

The main remaining task is to describe the physical properties of
the gauged height model (\ref{g7}); we will embark on this in
Section~\ref{gh}. However, before we do so, we will,
for completeness, present the analog of the compact
QED models of Section~\ref{cqed}.

\subsubsection{Compact QED}
\label{cqed2}

We apply the method of Section~\ref{cqed} to $\widetilde{Z}_H$.
We define integer-valued fields $M_{a \mu} = \Delta_{\mu} N_a -2 a_{a \mu}$
and $b_{i \mu} = \epsilon_{\mu\nu\lambda} \Delta_{\nu} a_{a\lambda}$
which satisfy the constraints $\epsilon_{\mu\nu\lambda} \Delta_{\nu} M_{a \lambda}
+ 2 b_{i \mu} = 0$ and $\Delta_{\mu} b_{i \mu} = 0$. Imposing
these constraints by continuous real fields $A_{i \mu}$ and $\phi_i$
respectively, and performing the summation over $M_{a \mu}$
and $b_{i \mu}$ by the Poisson summation formula we obtain
\begin{eqnarray}
\widetilde{Z}_{cQED}= && \sum_{\{Q_{a\mu}\},\{p_{i \mu}\}} \prod_{i,\mu}
\int_{-\pi}^{\pi} \frac{d A_{i \mu}}{2 \pi}
\prod_{i}
\int_{-\pi}^{\pi} \frac{d \phi_{i}}{2 \pi}
\exp \left(
- \frac{1}{2 e^2} \sum_{i,\mu} \left( \epsilon_{\mu\nu\lambda} \Delta_{\nu}
A_{i \lambda} - 2 \pi Q_{a \mu} \right)^2  - \right.\nonumber \\
&&~~~~~~~~~~~~~~~~~~~~~~~~
\left. \frac{1}{2 g} \sum_{i,\mu} \left(
\Delta_{\mu} \phi_{i \mu} - 2 A_{i \mu} - 2 \pi p_{i\mu} \right)^2
+ i 2S \sum_i \eta_i A_{i\tau} \right),
\label{g10}
\end{eqnarray}
where $Q_{a\mu}$ and $p_{i \mu}$ extend over all the integers.
This is the Villain form of the action of compact QED coupled to a
charge 2 Higgs scalar\cite{rsprl1,srijmp}.
The Berry phase term,
attributable to the background charge density of $2S \eta_i $
on each site, is the same as that in $Z_{cQED}$ in (\ref{f10a}).
If, as in Section~\ref{cqed}, we had applied the
transformation of this subsection to $\widetilde{Z}_H^{\prime}$
rather than $\widetilde{Z}_{H}$, then the Berry phases would have
been attached to the monopoles and vortices of the compact QED
theory\cite{js2}; this is straightforward to do and we will not explicitly
present the results here---the final form is, of course, exactly
equivalent to (\ref{g10}), (\ref{g8}) or (\ref{g7}).
Also note that the $g=\infty$ limit of $\widetilde{Z}_{cQED}$
in (\ref{g10})
is exactly equal to $Z_{cQED}$ in (\ref{f10a}).

\subsubsection{Gauged height models}
\label{gh}
We now return to the main objective of this section: determination
of the phase diagram of the gauged height model $\widetilde{Z}_H$
in (\ref{g7}). We will restrict our attention to $S=1/2$; results
for other half-odd-integer values of $S$ are similar, and the
generalization to integer $S$ is straightforward.

\begin{figure}
\centering
\includegraphics[width=4in]{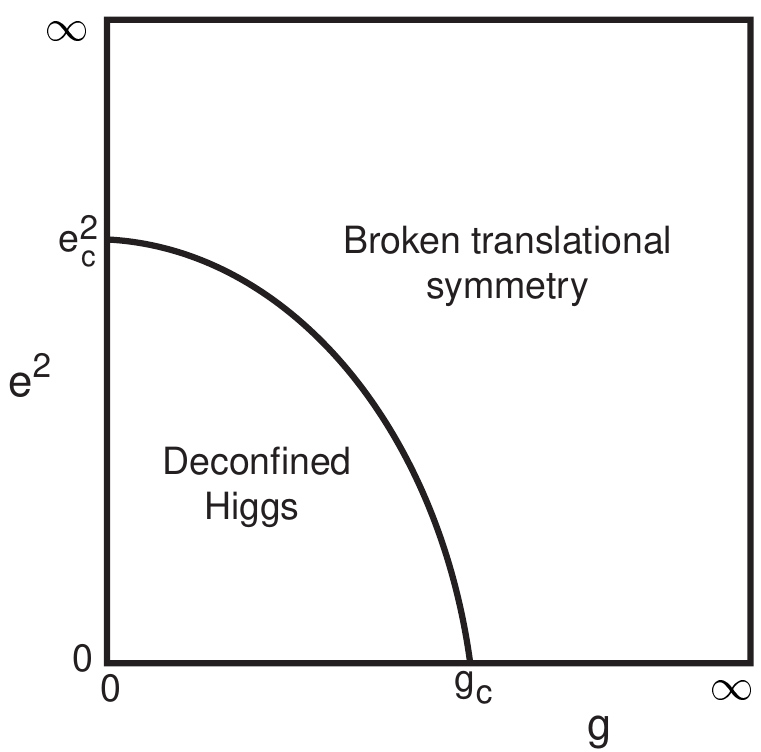}
\caption{
Phase diagram of the gauged height model $\widetilde{Z}_H$ in
(\protect\ref{g7}) for $S=1/2$; the same phase diagram also applies to the
height model $\widetilde{Z}_H^{\prime}$
in (\protect\ref{g8}), or to the exactly equivalent
compact QED + charge 2 Higgs scalar
theory $\widetilde{Z}_{cQED}$ in (\protect\ref{g10}).
There is no translational symmetry breaking in the `deconfined
Higgs' phase. The phase with translational symmetry breaking is
 confining; however, it is not completely ruled out that a
portion of it could be deconfining, although no such deconfined
phase with broken translational symmetry appears in
Refs~\protect\cite{js2,bmb,grest}. Most of the translational
symmetry breaking is in the columnar pattern of
Fig~\protect\ref{fig1}a; however for very large $e^2$ there
appears to be~\protect\cite{grest} a change by a first order
transition into the plaquette state of
Fig~\protect\ref{fig4}--this is not shown. The entire phase
transition line bounding the deconfined, Higgs phase
(apart from the single point $e^2 = 0$, $g=g_c$) is described
by the fully-frustrated Ising model in a transverse field,
$Z_I$ in (\protect\ref{g12}): this
transition is expected to be in the $D=3$ XY universality class,
with the {\em high} temperature phase of the XY model corresponding
to the Higgs phase. The transition at $e^2 =0$, $g=g_c$ is also in
the $D=3$ XY universality class, but it is inverted with respect
to the previous one---now the {\em low} temperature phase of the XY
model is the Higgs phase.
}
\label{fig6}
\end{figure}

As in the work by Fradkin and Shenker \cite{fradshen},
it is useful to look at various limiting values of the couplings
in $\widetilde{Z}_H$: from these results we have constructed the
phase diagram shown in Fig~\ref{fig6}.\\
({\em i\/}) $g=\infty$: \\
The field $a_{a\mu}$ must be `pure gauge'
along this line, with $a_{a\mu} = \Delta_\mu c_{a}$ for some
integer-valued $c_a$. This can be absorbed in $N_a$ by
$N_a \rightarrow N_a + 2 c_a$, and so $\widetilde{Z}_H$ reduces to
the pure height model $Z_H$ considered in (\ref{e3a}) in
Section~\ref{bm}. So the heights are in a smooth phase, the gauge
theory is confining, and there is translational symmetry breaking
due to the presence of bond-charge-density order.\\
({\em ii\/}) $e^2 = 0$:\\
Now the height field $N_a$ drops out, and the theory is simply
that of the integer-valued gauge field $a_{a\mu}$ with a Maxwell
action. This theory is exactly equivalent to the pure XY model in
2+1 dimensions, as can be seen by setting $A_{i\mu} = Q_{a\mu} = 0$
in (\ref{g10}). So there is a phase transition in the D=3 XY
universality class at a critical $g=g_c$; the $g<g_c$ ($g>g_c$)
phase maps onto the low (high) temperature phase of the XY
model.\\
({\em iii\/}) $g=0$:\\
This is the most interesting limit, where new physics emerges.
With no Maxwell term for the gauge field $a_{a\mu}$, the sum over $a_{a\mu}$
can be performed independently on each link. From this it is
evident that only the {\em parity} of the $N_a$ is relevant--the
partition function depends only upon whether a given $N_a$ is even
or odd. So we can introduce an Ising spin variable, $\sigma_a$ on each site of the
dual lattice by
defining
\begin{equation}
\sigma_a = 2 \left[ N_a (\mbox{mod 2}) \right] -1.
\label{g11}
\end{equation}
In terms of the $\sigma_a$, $\widetilde{Z}_H$ reduces to the Ising
partition function in three dimensions\cite{js2}
\begin{equation}
Z_I = \sum_{\{\sigma_a=\pm 1 \}} \exp\left( - \sum_{<a,b>} J_{ab}
\sigma_a \sigma_b \right).
\label{g12}
\end{equation}
The sum $a,b$ is over nearest neighbor pairs on the dual cubic
lattice. The exchange constants $J_{ab}$ all satisfy $|J_{ab}| =
J(e^2)$ where $J(e^2)$ is a monotonically increasing function defined by
\begin{equation}
e^{2 J(e^2)} \equiv \left[\sum_{n=-\infty}^{\infty} e^{-2 n^2
e^2}\right]\bigl/\left[\sum_{n=-\infty}^{\infty} e^{-(2n+1)^2 e^2 /2}\right].
\label{defj}
\end{equation}
The presence of the ${\cal X}_{a\mu}$ in $\widetilde{Z}_H$
introduces a key variation in the signs of the $J_{ab}$: it is not difficult to
see that these signs must be chosen so that
each plaquette in the $x$-$y$ plane is frustrated\cite{js2}. In the language
of 2+1 dimensional quantum models, this is the fully frustrated
Ising model in a transverse field\cite{isinggauge}. Such a model has been studied
in earlier works~\cite{js2,bmb,grest}, and we summarize the main results.
There is a phase transition at~\cite{js2,grest} $e^2 = e^2_c$
where $J(e_c^2) \approx
0.35$, such that for $e^2 > e^2_c$ both the Ising symmetry
($\langle \sigma_a \rangle \neq 0$) and the translational
symmetry are simultaneously broken. The broken translational
symmetry is due to development of columnar spin Peierls order of
the type shown in Fig~\ref{fig1}a. In the language of the gauge
theory, $\widetilde{Z}_{cQED}$, this phase is confining. For $e^2 <
e^2_c$, the gauge theory is in the deconfined `Higgs' phase, and Ising and
translational symmetries are restored.
This mechanism of deconfinement of spinons in antiferromagnets by
condensation of a charge 2 Higgs field was first proposed
in Ref \cite{rsprl2}. The condensation of the
Higgs field means that there are strong fluctuations
in the conjugate number operator {\em i.e.} in the number of
dimers carrying charges $\pm 2$---these can fluctuate into dimers
of net charge 0, without violating the number constraint which
fixes the total number of singlet bonds from each site at $\pm
2S$, as shown in Fig~\ref{fig5}.
According to the arguments
of Refs~\cite{js2,bmb} the phase transition in the fully
frustrated Ising model at $e^2 = e^2_c$ is also in the XY
universality class, but now the $e^2 > e^2_c$ ($e^2 < e^2_c$)
region maps onto the low (high) temperature phase of the XY model.
Finally we note that the simulations of Grest~\cite{grest} appear to observe a
second phase transition at $e^2 = e^2_{2c}$ where $ J(e_{2c}^2) \approx 1.8$:
for $e^2 > e^2_{2c}$
the pattern of the translational symmetry breaking appears to
change into the plaquette phase of Fig~\ref{fig4}---however the
gauge theory remains in the confining phase.\\
({\em iv\/}) $e^2 = \infty$:\\
The physics here is very similar to the large $e^2$ region
considered in ({\em iii\/}) at $g=0$---translational and Ising
symmetries are broken and the gauge theory is confining.

All of the above results have been combined in Fig~\ref{fig6}
which is one of the central results of this paper.
There is a
single phase boundary connecting the phase transitions at
$e^2 = 0$, $g=g_c$ and $e^2 = e^2_c$, $g=0$ found above. By
familiar arguments~\cite{fradshen} it is expected that the
universality class of the transition along this entire line will
be the same as that of the $e^2 = e^2_c$, $g=0$ point, and the
$e^2 = 0$, $g=g_c$ point is a singular limit.

\section{Doped Antiferromagnets}
\label{dope}

We now discuss recent work\cite{vojta} which examined the consequences of
hole-doping the state with broken translational symmetry in
Fig~\ref{fig6}. A theory for hole-doping the
deconfined Higgs phase of Fig~\ref{fig6} does not exist and would
be an interesting direction for future work.

The study was carried out in a particular large $N$ limit in which
the columnar spin-Peierls state of Fig~\ref{fig1}a
was a ground state in mean-field theory. Further, the ordering in
the state was fully developed, and the ground state
was well separated from any
possible
phase transition to a deconfined Higgs phase or to a magnetically
ordered state. This is probably not a realistic situation for the
insulator, but it is expected that the results so obtained are
generic for doping an arbitrary spin-Peierls phase. The results of
the study have also been reviewed recently\cite{lt}, and so we will limit
ourselves here to highlighting the main results and mentioning
some experimental implications.

\begin{figure}
\centering
\includegraphics[width=5.5in]{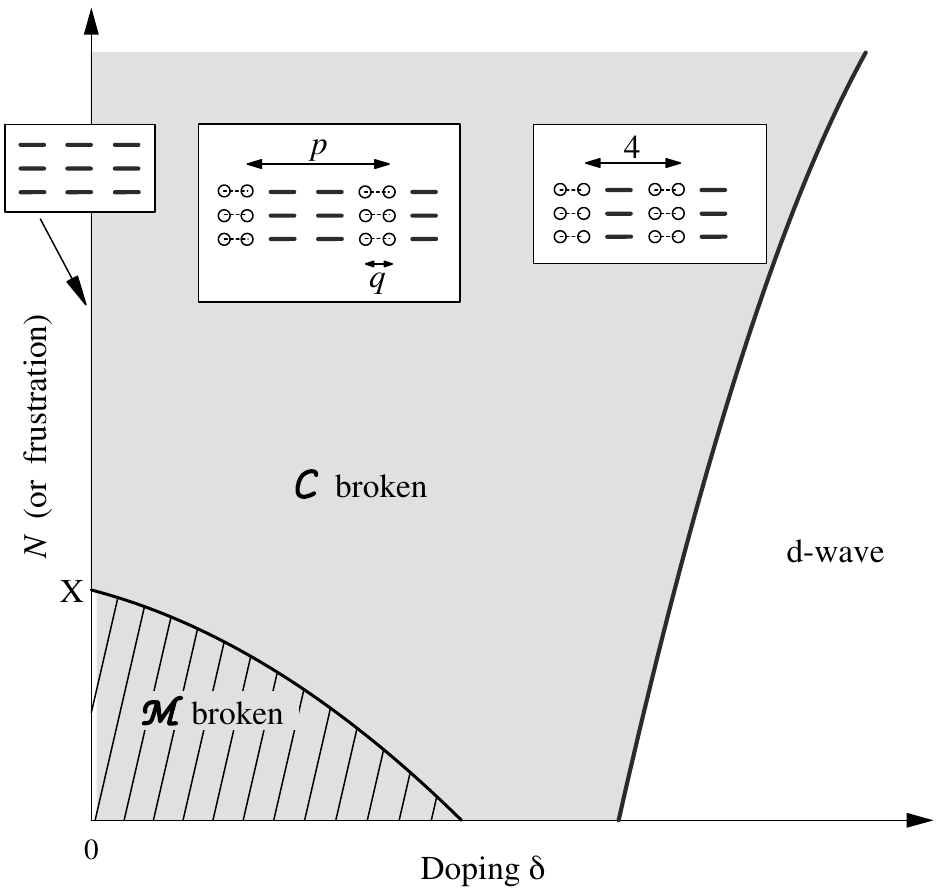}
\caption{
Ground state phase diagram of a doped antiferromagnet, adapted from
Ref~\protect\cite{vojta}. ${\cal M}$ is the symmetry of spin rotations
and it is broken in the
hatched region, while ${\cal C}$ is the translational symmetry which broken
in the shaded region.
At zero doping, the physical system has N\'{e}el order and so ${\cal M}$
is broken. By using some frustration in the spin Hamiltonian (or
considering models with a larger spin symmetry group (large $N$)),
we move the system across the quantum critical point $X$ into
state with columnar spin-Peierls (bond-charge-density) order.
The states obtained by doping this state are sketched:
the thick and dashed lines indicate varying
values of the bond charge density, while the circles represent site hole density.
All the states with non-zero hole density are superconducting for
large enough $N$, but could undergo phase transitions to
insulating, anisotropic Wigner crystalline states as $N$ is
reduced.
}
\label{fig7}
\end{figure}

As in Refs~\cite{zaanen,schulz}, the main effect found was that
for reasonable values of the microscopic parameters, the holes
prefer to segregate in one-dimensional striped structures with a
finite density per unit length---so as the net hole concentration
tends to zero, the stripes move farther apart but maintain a
finite hole density per unit length per stripe. However, the
existence of a background of spin-Peierls order makes the details
of the hole configurations quite different from earlier
work~\cite{zaanen,schulz,ekl}
in experimentally significant ways, as we detail below.
A schematic for the evolution
of the hole configuration is indicated in the phase diagram in
Fig~\ref{fig7}.

The unit cells of the charge ordered states in Fig~\ref{fig7} have
size $p \times 1$; within each unit cell, the holes are
concentrated in a region of size $q \times 1$. Both $p$ and $q$
were always found to be even, with $q$ remaining fixed while $p \rightarrow \infty$
as the doping concentration $\delta \rightarrow 0$. Significant
properties of these even-width stripes are:
\begin{itemize}
\item
The holes density per unit length in the $q$-width region is not
unity (as found in earlier theories~\cite{zaanen,schulz})
but various continuously dependent upon microscopic parameters.
For reasonable choices we can obtain a hole density per unit
length of around 0.5 (for $q=2$ this corresponds to a ladder with $\approx 1/4$
hole per site), but it is never pinned at exactly this value
for any finite range of doping, $\delta$ {\em i.e.} the stripes
are not incompressible.
\item
There are strong pairing correlations between the holes in each
$q$-width region. This leads to an anisotropic superconducting
ground state, and should allow good metallic conduction above the
superconducting transition temperature. These characteristics are
consistent with observations\cite{uchida} on ${\rm La}_{2-x-y}{\rm Nd}_y {\rm
Sr}_x {\rm Cu O}_4$.
\item
The evolution of the charge ordering wavevector ($K=1/p$)
as a function of doping, $\delta$,
is shown in Fig~\ref{fig8}.
\begin{figure}
\centering
\includegraphics[width=3.5in]{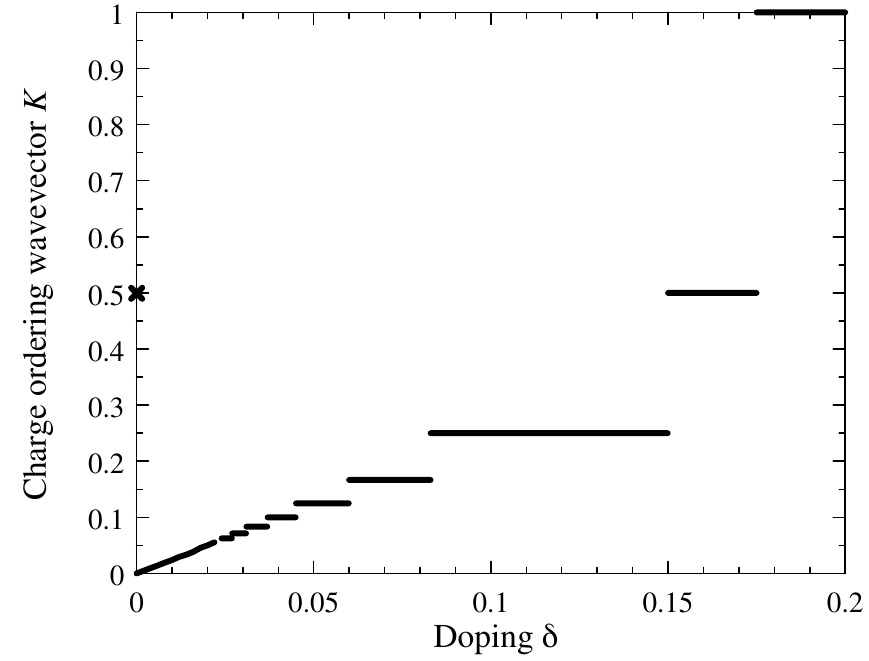}
\caption{
Charge ordering wavevector ($1/p$) as a function of the doping, $\delta$
(from Ref~\protect\cite{vojta}).}
\label{fig8}
\end{figure}
Note that this wavevector is always quantized at values which
equal 1/(even integer). The plateaus are quite small for small
$\delta$, but become progressively broader as the even integer
decreases. In particular, as is observed experimentally, a large
plateau at wavevector 1/4 is present and this emerges naturally
from our theory--we regard this as significant.
Plateaus at smaller wavevector (say 1/6) have not been
experimentally observed, but the resolution of current experiments
is not precise enough to distinguish the staircase-like evolution
in Fig~\ref{fig8} from a continuous variation before reaching the
plateau at 1/4.
The hole density
per unit length of stripe varies continuously within each plateau
but jumps discontinuously as the transition is made from one
plateau to the next.
\item
Note that, in Fig~\ref{fig8}, before reaching the d-wave superconducting state with
$p=1$, there is also a small plateau at $p=2$:
this is a state with uniform site-charge density, but with the
bond-charge density of Fig~\ref{fig1}a coexisting with d-wave-like
superconductivity. Such a state has not been observed, and it
would be interesting to look for one.
\item All of the stripes found in this computations are
bond-centered, at least in the region where
the symmetry of spin-rotations, ${\cal M}$, remains unbroken;
we cannot rule out the possibility that there will be a transition
to site-centered stripes, like those found earlier~\cite{zaanen,schulz},
once ${\cal M}$ is broken.
It would be useful for
future experiments to detect this distinction
between bond-centered and site-centered stripes. Let us discuss this
difference more precisely for the case $p=4$. For the site-centered
stripe, the hole density per
unit length in each column of sites takes values $\rho_1$,
$\rho_2$, $\rho_3$, $\rho_2$ before repeating periodically (where
$\rho_{1-3}$ are some three distinct densities); this is the configuration
usually assumed in most experimental papers. In contrast for the
bond-centered state, these densities take the values
$\rho_1$, $\rho_1$, $\rho_2$, $\rho_2$, and this also appears to
be compatible with existing observations. In the computation here
the bond-centering is important for enhancing pairing
correlations, which are responsible for superconducting/metallic
transport in the direction parallel to the stripes.
We note that a very recent NQR experiment\cite{julich} indicates
a charge distribution which has some features
consistent with the bond-centered state: they find only two
inequivalent sites associated with the stripes, and a density per
site in
the hole-rich region which is only about 0.18-0.19 (the fully
segregated site-centered stripe would have $\rho_1 \approx 0.5$, $\rho_2=\rho_3
=0$,
while the fully segregated bond-centered stripe would have $\rho_1
\approx 0.25$, $\rho_2 = 0$; clearly the bond-centered value is more compatible
with observations).
\end{itemize}

\section*{Acknowledgements}
We thank G.~Teitel'baum, S.~Uchida, Jan~Zaanen, and especially Matthew Fisher and
T.~Senthil
for useful discussions. We are particularly grateful to
the last two for describing
their results prior to publication.
S.S. thanks Masashi Takigawa for organizing a stimulating conference,
and wishes Prof. Yasuoka many more productive years.
This research was supported by
US NSF Grant No DMR 96--23181 and by the DFG (VO 794/1-1).

\newpage 

\foreach \x in {1,...,9}
{%
\clearpage
\includepdf[pages={\x}]{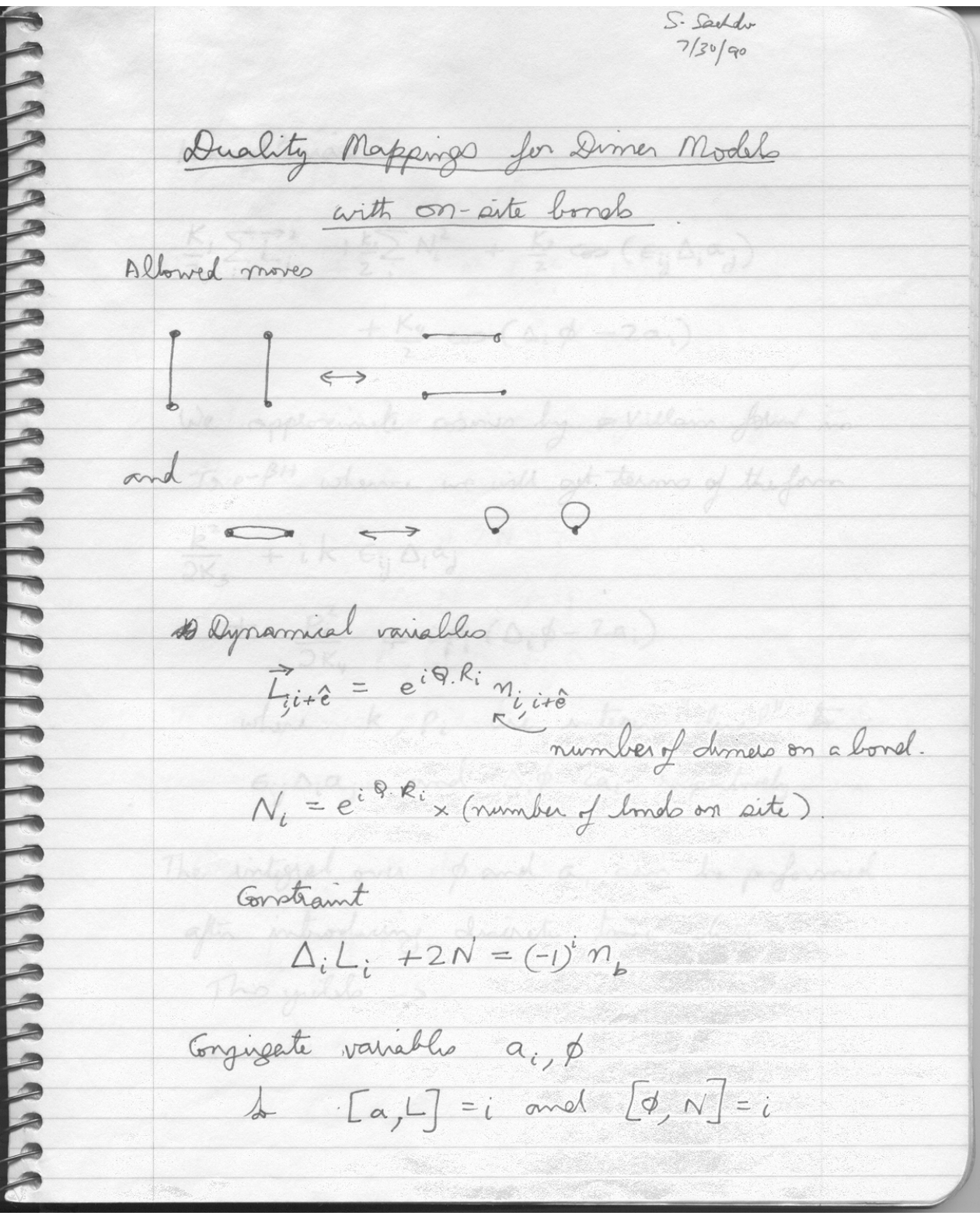} 
}


\begin{thebibliography}{99}

\bibitem{jtran}
J.~M.~Tranquada, J. Phys. Chem. Solids {\bf 59}, 2150 (1998).

\bibitem{birg} S.~Wakimoto {\em et al}, Phys. Rev. B {\bf 60},
R769 (1999); Y.~S.~Lee {\em et al}, {\em ibid} {\bf 60}, 3643
(1999).

\bibitem{imai} A.~W.~Hunt, P.~M.~Singer, K.~R.~Thurber, and T.~Imai,
Phys. Rev. Lett. {\bf 82},
4300 (1999);
T.~Imai, C.~P.~Slichter, K.~Yoshimura,
and K.~Kosuge, Phys. Rev. Lett. {\bf 70}, 1002 (1993);
T.~Imai, C.~P.~Slichter, K.~Yoshimura,
M.~Katoh,
and K.~Kosuge, Phys. Rev. Lett. {\bf 71}, 1254 (1993);
S.~Fujiyama
M.~Takigawa, Y.~Ueda, T.~Suzuki, and N.~Yamada,
cond-mat/9904275.

\bibitem{rsprl1} N.~Read and S.~Sachdev, Phys. Rev. Lett. {\bf 62},
1694 (1989).

\bibitem{rsprb1} N.~Read and S.~Sachdev, Phys. Rev. B {\bf 42}, 4568 (1990).


\bibitem{zaanen} J.~Zaanen and O.~Gunnarsson, Phys. Rev. B {\bf 40},
7391 (1989).

\bibitem{schulz} H.~Schulz, J. de Physique {\bf 50}, 2833 (1989).

\bibitem{vojta} M.~Vojta and S.~Sachdev, Phys. Rev. Lett. {\bf 83},
Nov 8 (1999); cond-mat/9906104.

\bibitem{lt} S.~Sachdev and M.~Vojta, cond-mat/9908008.

\bibitem{dimer} D.~Rokhsar and S.~A.~Kivelson, Phys. Rev. Lett. {\bf 61}, 2376
(1988).

\bibitem{rsnpb} N.~Read and S.~Sachdev, Nucl. Phys. B {\bf 316},
609 (1989).

\bibitem{fradkiv} E.~Fradkin and S.~Kivelson, Mod. Phys. Lett. B
{\bf 4}, 225 (1990).

\bibitem{js1} R.~A.~Jalabert and S.~Sachdev, Mod. Phys. Lett. B {\bf 4}, 1043
(1990).

\bibitem{js2} R.~A.~Jalabert and S.~Sachdev, Phys. Rev. B {\bf 44},
686 (1991).

\bibitem{ss} S.~Sachdev, unpublished notes, July 30, 1990.

\bibitem{senthil} T.~Senthil and M.~P.~A.~Fisher,
cond-mat/9910224.

\bibitem{nagaosa} N.~Nagaosa and P.~A.~Lee, cond-mat/9907019.

\bibitem{zheng} W.~Zheng and S.~Sachdev, Phys. Rev. B {\bf 40}, 2704 (1989).

\bibitem{haldane} F.~D.~M.~Haldane, Phys. Rev. Lett.
{\bf 61}, 1029 (1988).

\bibitem{cavo} S. Sachdev and N. Read, Phys. Rev. Lett. {\bf 77}, 4800
(1996), footnote 19.

\bibitem{rsprl2} N.~Read and S.~Sachdev, Phys. Rev. Lett. {\bf 66},
1773 (1991).

\bibitem{srijmp} S.~Sachdev and N.~Read, Int. J. Mod. Phys. B
{\bf 5}, 219 (1991).

\bibitem{fradshen} E.~Fradkin and S.~H.~Shenker, Phys. Rev. D {\bf
19}, 3682 (1979).

\bibitem{isinggauge} It is also instructive to take the $g=0$
limit of the action, $\widetilde{Z}_{cQED}$, for
compact QED coupled to a charge 2 Higgs scalar in
(\protect\ref{g10}). Then in a suitable gauge
$A_{i\mu}$ is pinned to 0 or $\pi$. We can now define an Ising
{\em gauge} field $\zeta_{i\mu} = e^{iA_{i\mu}}$, and the action in
(\protect\ref{g10}) reduces to the standard plaquette action for a
2+1 dimensional $Z_2$ gauge theory on the direct lattice.
The Berry phase term
contributes a factor $\prod_{i} \zeta_{i\tau}^{2S}$ to the partition
function, and this is the
representation of the frustration in the
dual lattice Ising model $Z_I$ in (\protect\ref{g12}).


\bibitem{bmb} D.~Blankschtein, M.~Ma, and A.~N.~Berker, Phys. Rev. B
{\bf 30}, 1362 (1984).

\bibitem{grest} G.~S.~Grest, J. Phys. C: Solid State Phys.
{\bf 18}, 6239 (1985).

\bibitem{ekl} V.~J.~Emery, S.~A.~Kivelson, and H.~Q.~Lin, Phys.
Rev. Lett. {\bf 64}, 475 (1990);
S.~A.~Kivelson, V.~J.~Emery, and H.~Q.~Lin,
Phys. Rev. B {\bf 42}, 6523 (1990).

\bibitem{uchida} N.~Ichikawa, S.~Uchida, J.~M.~Tranquada,
T.~Niemoeller, P.~M.~Gehring, S.-H.~Lee, and J.~R.~Schneider,
cond-mat/9910037.

\bibitem{julich} G.~B.~Teitel'baum, B.~B\"uchner, and
H.~de~Gronckel, cond-mat/9910036.

\end{thebibliography}
\end{document}